\documentclass[conference]{IEEEtran}
\IEEEoverridecommandlockouts

\usepackage{cite}
\usepackage{amsmath,amssymb,amsfonts}
\usepackage{algorithmic}
\usepackage{graphicx}
\usepackage{textcomp}
\usepackage{xcolor}
\def\BibTeX{{\rm B\kern-.05em{\sc i\kern-.025em b}\kern-.08em
    T\kern-.1667em\lower.7ex\hbox{E}\kern-.125emX}}

\usepackage{amsfonts}
\usepackage{booktabs}
\usepackage{siunitx}
\usepackage{makecell}
\usepackage{multirow}
\usepackage{xcolor}
\usepackage[utf8]{inputenc}
\usepackage{url}
\usepackage{tabularx}
\usepackage{threeparttable}
\usepackage{array}
\usepackage{algorithm}
\usepackage{algorithmic}

\usepackage{etoolbox}

\usepackage{amsmath,amssymb}

\newcommand{\MASK}{\text{\ttfamily [MASK]}}
\newcommand{\EOS}{\text{\ttfamily [EOS]}}

\definecolor{metricgray}{gray}{0.45}
\newcolumntype{L}[1]{>{\raggedright\arraybackslash}p{#1}}

\newcommand{\metric}[1]{#1}
\newcommand{\metricpm}[2]{#1{\scriptsize\textcolor{metricgray}{\,\ensuremath{\pm}#2}}}

\newcommand{\dciinline}[2]{#1{\scriptsize\textcolor{metricgray}{\,\ensuremath{[#2]}}}}

\definecolor{metricgray}{gray}{0.45}
\newcolumntype{L}[1]{>{\raggedright\arraybackslash}p{#1}}

\begin{document}



\title{Mask, Sample, Revise: A Revisable CTMC Inference Stack for Guided Discrete Flow Matching Text-to-Speech}

\author{
\IEEEauthorblockN{
Alef Iury Ferreira\textsuperscript{1},
Lucas Rafael Gris\textsuperscript{1},
Frederico Oliveira\textsuperscript{2},
Christopher Dane Shulby\textsuperscript{3},\\
Luiz Fernando Vidal\textsuperscript{1},
Anderson Soares\textsuperscript{1},
Arlindo Galv\~ao Filho\textsuperscript{1}
}
\vspace{3pt}
\IEEEauthorblockA{
\textsuperscript{1}\textit{Federal University of Goi\'as}, Goi\^ania, Brazil\\
\textsuperscript{2}\textit{Federal University of Mato Grosso}, Cuiab\'a, Brazil\\
\textsuperscript{3}\textit{Elsa Speak}, S\~ao Paulo, Brazil\\[4pt]
alef\_iury\_c.c@discente.ufg.br
}
}

\definecolor{pfgblue}{RGB}{31,119,180}
\definecolor{remaskorange}{RGB}{214,97,0}

\newcommand{\pfg}[1]{\textcolor{pfgblue}{#1}}
\newcommand{\remask}[1]{\textcolor{remaskorange}{#1}}
\newcommand{\algcmt}[1]{\hfill{\scriptsize\textcolor{gray}{// #1}}}

\makeatletter
\newcommand{\acceptednotice}{%
  Accepted at the 2026 IEEE Spoken Language Technology Workshop (SLT 2026)}
\newcommand{\copyrightnotice}{%
  \parbox{\textwidth}{\centering\scriptsize
  \copyright\,2026 IEEE. Personal use of this material is permitted. Permission from IEEE must be obtained for all other uses, in any current or future media, including reprinting/republishing this material for advertising or promotional purposes, creating new collective works, for resale or redistribution to servers or lists, or reuse of any copyrighted component of this work in other works.}}
\def\ps@IEEEtitlepagestyle{%
  \def\@oddhead{\hfill\footnotesize\acceptednotice\hfill}%
  \def\@evenhead{}%
  \def\@oddfoot{\hfill\copyrightnotice\hfill}%
  \def\@evenfoot{}%
}
\makeatother

\maketitle
\thispagestyle{IEEEtitlepagestyle}

\maketitle

\begin{abstract}

Recent alignment-free non-autoregressive (NAR) text-to-speech (TTS) models formulate synthesis as a conditional infilling task, bypassing explicit duration predictors and external aligners. When speech is represented with neural codec tokens, the infilling problem becomes discrete, making Discrete Flow Matching (DFM), a Continuous-Time Markov Chain (CTMC) framework for discrete generation, a natural fit. However, inference-time control for stable low-step conditional infilling remains underexplored. We propose Mask, Sample, Revise, an inference-time CTMC stack for alignment-free DFM-TTS. The stack combines predictor-free guidance to strengthen text conditioning, prompt-matched conditional coupling to align the probability path with the acoustic prompt, and SC-ReMask, a schedule-constrained remasking mechanism that introduces token-to-mask transitions so de-masking decisions can be revised. These components require no post-hoc fine-tuning and operate in a single tau-leaping sampler. Controlled ablations show that this stack improves intelligibility and robustness in the low-NFE prompted setting, outperforming unguided and guidance-only samplers with substantially more steps.

\end{abstract}


\section{Introduction}
\label{sec:introduction}

Modern text-to-speech (TTS) systems can be broadly divided into two paradigms: autoregressive (AR) and non-autoregressive (NAR). AR models like VALL-E~\cite{wang2023neural} offer strong temporal modeling but suffer from slow inference, while NAR models enable parallel generation at the cost of often requiring external aligners and duration predictors that can constrain prosody and increase complexity~\cite{chen-etal-2025-f5}. A promising NAR direction is alignment-free infilling, where synthesis is cast as a conditional infilling problem over acoustic representations. For example, E2-TTS~\cite{eskimez2024e2} uses a ``text-filler'' strategy that pads the text representation to match the target acoustic length and trains the model to infill masked regions, removing the need for explicit duration modules. However, alignment-free infilling can be brittle at inference time: conditional control may weaken at low sampling budgets, and early mistakes can persist as deletions, substitutions, or speaker drift~\cite{chen-etal-2025-f5}.

Concurrently, Discrete Flow Matching (DFM)~\cite{gat2024discrete} has emerged as a principled framework for discrete generation, formulated as a Continuous-Time Markov Chain (CTMC) where a learned transition-rate field transports a simple source distribution, often a fully masked sequence, toward the data distribution. This is naturally suited to neural codec-based TTS, where speech is represented as a discrete sequence. Recent systems such as DiFlow-TTS~\cite{nguyen2025diflow}, H-DFM~\cite{lee2026hierarchical}, and GibbsTTS~\cite{yang2026kinetic} demonstrate the promise of DFM for codec-token TTS, emphasizing factorized representations, Residual Vector Quantization (RVQ) aware modeling, or probability-path scheduling. Yet, it remains unclear which inference-time controls are necessary for stable conditional sampling in alignment-free mask-source infilling settings. In parallel, Discrete Guidance~\cite{nisonoff2025unlocking} derives principled guidance rules for CTMCs, showing that inference-time rate control can substantially strengthen conditional generation.

In this work, we study the inference procedure itself: how alignment-free codec-token TTS can benefit from combining sampling-time mechanisms within a DFM formulation. We propose a unified inference control stack comprising: (1) predictor-free guidance (PFG), which blends CTMC transition rates to strengthen conditioning; (2) conditional coupling, which constructs conditional probability paths under prompting; and (3) a remasking mechanism that injects token-to-mask transitions during sampling to correct early errors, which we call SC-ReMask: Schedule-Constrained CTMC Remasking. These components require no post-hoc fine-tuning. An important advantage of the discrete CTMC/DFM setting is that generation can be made revisable through explicit token-to-mask transitions during sampling. In contrast, systems such as E2-TTS~\cite{eskimez2024e2} and F5-TTS~\cite{chen-etal-2025-f5} operate in a continuous acoustic-feature setting, where this form of discrete remasking is not directly available. To the best of our knowledge, this is the first work to apply CTMC discrete guidance to TTS, introduce revisable remasking for alignment-free TTS, and combine PFG, prompt-matched conditional coupling, and token-to-mask remasking within a single alignment-free codec-token DFM framework. We refer to the resulting system as G-DFlow-TTS, whose central component is the Mask, Sample, Revise inference stack. We evaluate with objective metrics and human listening tests including paired statistical testing, and provide a demo page\footnote{\url{https://gdflowtts.github.io/G-DFlowTTS-Demo}} and source code\footnote{\url{https://github.com/alefiury/G-DFlowTTS}}.

\noindent\textbf{Contributions.} Our main contributions are:
\begin{itemize}

    \item We propose SC-ReMask, a remasking mechanism adapted from masked discrete diffusion to DFM by implementing token-to-mask moves as explicit CTMC transitions. Making discrete infilling revisable during generation.

    \item We introduce a revisable CTMC inference stack for DFM-TTS, combining predictor-free guidance, prompt-matched conditional coupling, and schedule-constrained remasking within a single tau-leaping sampler.

    \item We provide controlled ablations showing that inference-time control, rather than merely increasing the number of sampling steps, is the main factor behind improved intelligibility in the prompted low-NFE setting.
\end{itemize}

\section{Related Work}
\label{sec:related_work}

NAR TTS has increasingly adopted powerful generative frameworks such as diffusion~\cite{ho2020denoising} and flow matching~\cite{lipman2023flow}. State-of-the-art (SOTA) systems, including NaturalSpeech~3~\cite{junaturalspeech}, VoiceBox~\cite{le2023voicebox}, and Matcha-TTS~\cite{mehta2024matcha}, demonstrate high-quality synthesis and efficient sampling, but many NAR pipelines still rely on explicit duration supervision or alignment modules to stabilize text-to-speech correspondence. While effective for intelligibility, such components can increase engineering complexity and may constrain prosodic variability~\cite{mayer25_interspeech}.

To improve flexibility, recent work has explored alignment-free strategies that reduce or eliminate explicit duration modeling. One line predicts coarse length information rather than phoneme-level durations. For example, DiTTo-TTS~\cite{lee2025dittotts} uses a Diffusion Transformer (DiT)~\cite{Peebles_2023_ICCV} conditioned on a predicted total sequence length to learn implicit alignment through cross-attention. Another strategy casts synthesis as an infilling task. E2-TTS~\cite{eskimez2024e2} popularized the ``text-filler'' approach, padding the text sequence with filler tokens to match the target acoustic length and training the model to infill masked regions. Subsequent works refine this paradigm with improved architectures and objectives, including F5-TTS~\cite{chen-etal-2025-f5} (continuous acoustic features) and MaskGCT~\cite{wang2025maskgct} (discrete token infilling). Although alignment-free infilling simplifies the pipeline, its robustness can depend heavily on the inference procedure, especially under small sampling budgets where early errors may persist~\cite{chen-etal-2025-f5}.

\begin{figure*}[htbp]
    \centering
    \includegraphics[width=\textwidth]{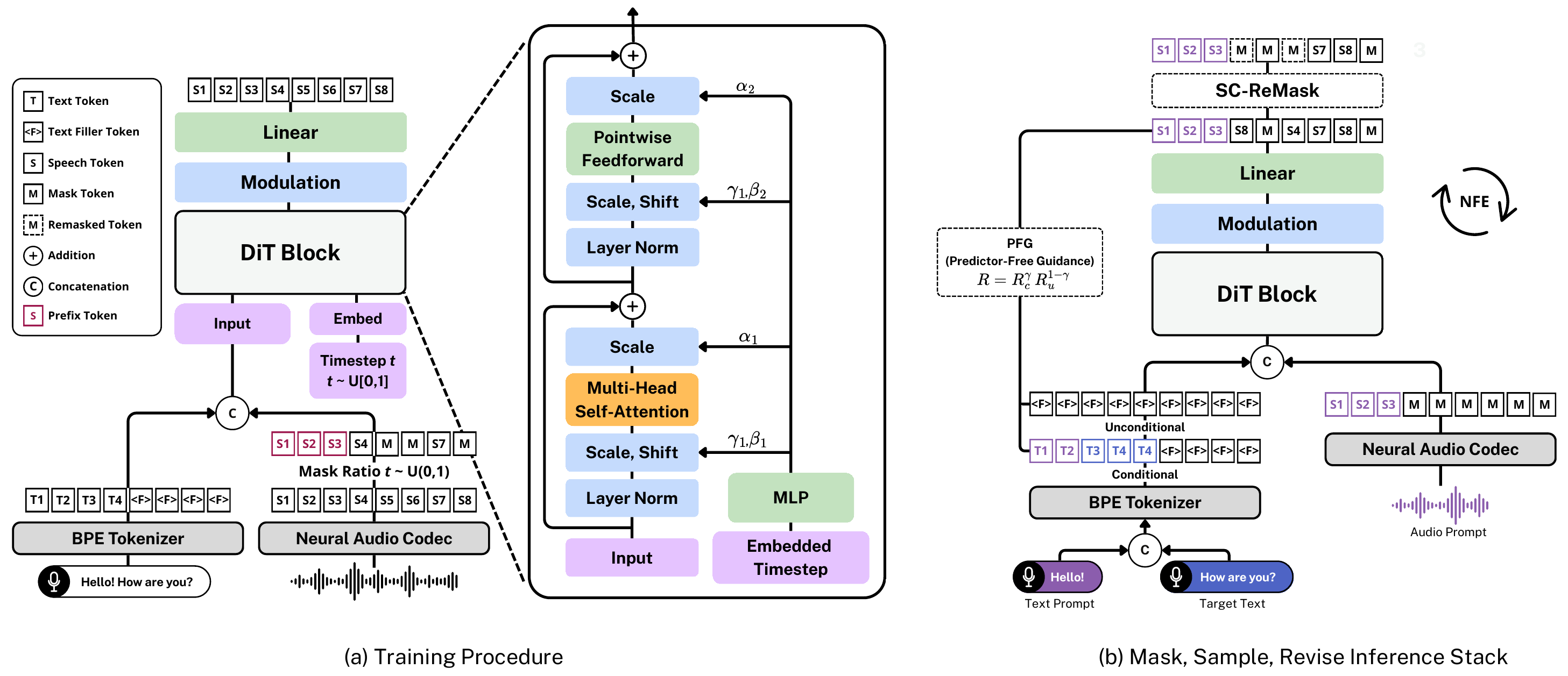}
    \caption{Overview of G-DFlow-TTS. (a) During training, the model learns DFM-based masked infilling over neural codec tokens using text-filler conditioning and prompt-matched conditional coupling. (b) At inference, the Mask, Sample, Revise stack combines predictor-free guidance (PFG), CTMC tau-leaping, and SC-ReMask token-to-mask transitions to revise generated tokens while keeping the acoustic prompt fixed.}
    \label{fig:architecture}
\end{figure*}


The move from continuous acoustic features to discrete speech tokens also connects TTS to broader discrete generative modeling for speech. Recent work by Ku et al.~\cite{ku2026discretediffusion} has explored this direction beyond TTS, studying discrete diffusion models (DDMs) for text-aligned speech token reconstruction by replacing an autoregressive decoder with a DDM-based decoder and analyzing sampler choices, inference steps, and remasking strategies. This work shows that inference-time sampling design is relevant for speech-token generation more broadly. However, the role of guided and revisable CTMC inference remains underexplored for alignment-free codec-token TTS, which is the setting addressed in this work.

Discrete Flow Matching (DFM)~\cite{gat2024discrete} provides a principled formulation of discrete generation via CTMC, learning transition-rate fields along probability paths between a source distribution, such as masked tokens, and the target data distribution. DFM has only recently been explored for TTS. DiFlow-TTS~\cite{nguyen2025diflow} studies compact and low-latency zero-shot TTS with factorized codec-token representations and dedicated prediction heads for different speech attributes. H-DFM~\cite{lee2026hierarchical} addresses the multi-codebook RVQ setting by aligning DFM training and inference with the codec hierarchy through coarse/fine modeling and a coarse-biased sampling schedule. GibbsTTS~\cite{yang2026kinetic} studies metric-induced DFM, focusing on kinetic-optimal scheduling~\cite{shaul2025flow} and finite-step moment correction for CTMC sampling. These works show that codec-token DFM is a promising direction, but they primarily emphasize representation design, codec hierarchy, probability paths, or solver correction.

These directions leave open the question of how to control the CTMC inference procedure itself in alignment-free DFM-TTS. In contrast to prior work, we focus on sampling-time control for mask-source codec-token generation. We build on discrete guidance for CTMC~\cite{nisonoff2025unlocking} and incorporate conditional coupling together with SC-ReMask, a schedule-constrained remasking mechanism inspired by ReMDM~\cite{wang2025remasking}. ReMDM shows that masked discrete generation can benefit from remasking already decoded tokens, enabling iterative refinement during inference. We adapt this idea to DFM by implementing token-to-mask remasking transitions as explicit CTMC transitions inside tau-leaping. This perspective targets stable conditional infilling through sampling-time rate control and makes discrete codec-token generation revisable during synthesis.

\section{G-DFlow-TTS: DFM Formulation and Revisable Inference Stack}
\label{sec:gdflowtts}

G-DFlow-TTS (Figure~\ref{fig:architecture}) denotes the resulting non-autoregressive TTS system built around the Mask, Sample, Revise inference stack. It synthesizes speech by infilling sequences of discrete neural audio codec tokens conditioned on text and an acoustic prompt. We use a DiT backbone and an alignment-free text-filler conditioning strategy, bypassing explicit duration predictors and external aligners. Because our goal is to study inference-time control, we keep the DFM-TTS architecture fixed and enhance the CTMC sampler with PFG, prompt-matched conditional coupling, and SC-ReMask.

Let $\mathcal{V}$ be the codec-token vocabulary, $m\in\mathcal{V}$ the mask token, and $\mathbf{x}\in\mathcal{V}^L$ a length-$L$ token sequence with positions indexed by $i$. DFM defines a time-indexed probability path $p_t$ over tokens for continuous time $t\in[0,1]$, using a monotone schedule $\kappa_t\in[0,1]$ with $\kappa_0=0$ and $\kappa_1=1$ (we use a polynomial convex scheduler $\kappa_t=t^{n}$ with $n=1$). With a masked source sequence $\mathbf{x}_0$ (all $m$) and a target sequence $\mathbf{x}_1$ (ground truth), we use the convex-mixture path from~\cite{gat2024discrete}:
\begin{equation}
\label{eq:convex_path}
p_t(x_i \mid \mathbf{x}_0, \mathbf{x}_1)
=
(1-\kappa_t)\,\delta_{x_{0,i}}(x_i) + \kappa_t\,\delta_{x_{1,i}}(x_i),
\end{equation}
where $\delta$ is the Kronecker delta. Training samples $t\sim\mathrm{Uniform}[0,1]$, draws $\mathbf{x}_t\sim p_t(\cdot\mid \mathbf{x}_0,\mathbf{x}_1)$, and minimizes cross-entropy to predict $\mathbf{x}_1$ from $(\mathbf{x}_t,y)$.

For conditional coupling, we modify the source sequence $\mathbf{x}_0$ by copying a random-length prefix from $\mathbf{x}_1$ and setting all remaining positions to $m$. The prefix duration is sampled uniformly in $[0.25, 12.0]$ seconds, ensuring at least $0.5$ seconds remain to be generated. The convex-mixture path (Eq.~\ref{eq:convex_path}) then operates over this coupled source.

At inference, we discretize into $K$ steps with $\Delta t = 1/K$ and sample via CTMC tau-leaping~\cite{gat2024discrete}.  In the base sampler, token-generation transitions are applied only to positions currently equal to $m$, yielding parallel unmasking over the suffix while keeping the prompt pinned. PFG is applied as a geometric mixture in rate space~\cite{nisonoff2025unlocking}: $R^{(\gamma)}_{i,v} = (R_{c,i,v})^{\gamma}\,(R_{u,i,v})^{1-\gamma}$, where $R_{c}$ and $R_{u}$ are the conditional and unconditional CTMC transition rates, $R_{i,v}\ge 0$ is the rate of changing position $i$ to token $v$, and $\gamma$ controls guidance strength.

\subsection{SC-ReMask}

SC-ReMask (Schedule-Constrained CTMC Remasking) makes discrete infilling revisable by adding token-to-mask transitions during inference. The method is inspired by ReMDM~\cite{wang2025remasking}, which introduces remasking for discrete diffusion through a per-step remask probability constrained by the noise schedule. We adapt this idea to DFM by implementing remasking as an explicit CTMC transition. As a result, remasking contributes directly to the total hazard and to the jump decisions made by tau-leaping, as summarized in Algorithm~\ref{alg:ctmc_remask}.

At inference step $k$, we first compute a schedule-constrained remask probability $\sigma(t_k)$. The switch time $t_{\mathrm{switch}}$ controls when remasking is enabled, allowing the sampler to postpone revision to selected stages of generation. Given consecutive times $t_k$ and $t_{k+1}$, we define the maximum admissible remasking probability as:

\begin{algorithm}[h!]
\caption{Guided CTMC inference with SC-ReMask. Blue lines denote PFG steps, while orange lines indicate SC-ReMask steps.}
\label{alg:ctmc_remask}
\footnotesize
\begin{algorithmic}[1]
\REQUIRE Text $y$, prompt codes $\mathbf{x}^{p}$, steps $K$, guidance $\gamma$
\STATE Initialize $\mathbf{x}\leftarrow[\mathbf{x}^{p},\MASK,\ldots,\MASK,\EOS]$
\FOR{$k=0,\ldots,K-1$}
\STATE $t_k\leftarrow k/K$, $\Delta t\leftarrow 1/K$
\STATE \pfg{$R_c\leftarrow R_\theta(\mathbf{x},y,t_k)$}, \quad \pfg{$R_u\leftarrow R_\theta(\mathbf{x},\varnothing,t_k)$} 
\STATE \pfg{$R\leftarrow R_c^{\gamma}R_u^{1-\gamma}$}
\STATE \remask{$\sigma\leftarrow\eta_{\mathrm{rescale}}\min(\eta_{\mathrm{cap}},\sigma_{\max}(t_k))$}
\STATE \remask{$\sigma\leftarrow 0$ if $t_k<t_{\mathrm{switch}}$}
\STATE \remask{$r^{\mathrm{rm}}\leftarrow-\log(1-\sigma)/\Delta t$}
\STATE \remask{Add rate $r^{\mathrm{rm}}$ for eligible token-to-$\MASK$ transitions}
\STATE $\mathbf{x}\leftarrow\mathrm{TauLeap}(\mathbf{x},R,\Delta t)$
\ENDFOR
\RETURN generated suffix of $\mathbf{x}$
\end{algorithmic}
\end{algorithm}

\begin{equation}
\label{eq:sigma_max}
\sigma_{\max}(t_k) = \min\left(1,\frac{1-\kappa_{t_{k+1}}}{\kappa_{t_k}}\right),
\end{equation}
with the convention that remasking is skipped when no suffix tokens are currently eligible. We then use the capped and rescaled schedule
\begin{equation}
\label{eq:sigma_schedule}
\sigma(t_k) = 
\begin{cases}
0, & t_k < t_{\mathrm{switch}}, \\
\eta_{\mathrm{rescale}} \min\left(\eta_{\mathrm{cap}},\sigma_{\max}(t_k)\right), & t_k \geq t_{\mathrm{switch}}.
\end{cases}
\end{equation}

Finally, we convert the per-step remasking probability into a CTMC rate by matching the tau-leap jump probability:
\begin{equation}
\label{eq:remask_rate}
r^{\mathrm{rm}}(t_k) = -\frac{\log(1-\sigma(t_k))}{\Delta t}.
\end{equation}

\begin{table*}[!ht]
\centering
\footnotesize
\caption{Results on LibriSpeech test-clean. We report mean scores with 95\% confidence intervals, where the gray $\pm$ values denote CI half-widths. External systems are shown using official checkpoints for contextual reference. $\dagger$ indicates a statistically significant improvement over the G-DFlow-TTS baseline at the same NFE (paired sign-flip permutation, $p<10^{-4}$; see Table~\ref{tab:significance}). Within the G-DFlow-TTS variants, the best result for each metric is highlighted in \textbf{bold}, and the second-best is \underline{underlined}.}
\label{tab:main_results_new}
\setlength{\tabcolsep}{3.2pt}
\renewcommand{\theadfont}{\bfseries}
\renewcommand{\arraystretch}{1.08}
\begin{tabular}{
    L{4cm}
    c
    c
    c
    c
    c
    c
    c
    c
}
\toprule
\thead{System} &
\thead{Params} &
\thead{Data} &
\thead{WER (\%) $\downarrow$} &
\thead{CER (\%) $\downarrow$} &
\thead{SIM-o $\uparrow$} &
\thead{UTMOS $\uparrow$} &
\thead{MOS $\uparrow$} &
\thead{RTF $\downarrow$} \\
\midrule

\multicolumn{9}{l}{\textbf{External Reference Systems}} \\
Ground Truth &
-- & -- &
\metricpm{2.29}{0.28} &
\metricpm{0.65}{0.09} &
\metricpm{0.76}{0.005} &
\metricpm{4.10}{0.02} &
\metricpm{4.08}{0.26} &
\metric{--} \\

MaskGCT~\cite{wang2025maskgct} &
1048M & 100K Multi. &
\metricpm{4.89}{0.40} &
\metricpm{1.90}{0.17} &
\metricpm{0.74}{0.003} &
\metricpm{3.88}{0.02} &
\metricpm{3.68}{0.27} &
\metric{--} \\

CosyVoice2~\cite{du2024cosyvoice} &
500M & 166K Multi. &
\metricpm{3.86}{0.33} &
\metricpm{1.47}{0.14} &
\metricpm{0.78}{0.003} &
\metricpm{4.35}{0.01} &
\metricpm{4.27}{0.22} &
\metric{--} \\

F5-TTS (32 NFE)~\cite{chen-etal-2025-f5} &
336M & 100K Multi. &
\metricpm{2.97}{0.32} &
\metricpm{0.88}{0.12} &
\metricpm{0.78}{0.003} &
\metricpm{3.90}{0.02} &
\metricpm{3.82}{0.24} &
\metric{--} \\
\midrule

\multicolumn{9}{l}{\textbf{Controlled G-DFlow-TTS Ablations}} \\
U-Coupling Baseline (32 NFE) &
232M & 60K EN &
\metricpm{75.44}{1.54} &
\metricpm{47.25}{1.07} &
\metricpm{0.17}{0.006} &
\metricpm{2.12}{0.03} &
\metric{--} &
\metric{0.05} \\

C-coupling only &
232M & 60K EN &
\metricpm{90.12}{1.47} &
\metricpm{57.79}{1.01} &
\metricpm{0.17}{0.007} &
\metricpm{1.80}{0.02} &
\metric{--} &
\metric{0.05} \\

U-coupling + PFG &
232M & 60K EN &
\metricpm{28.61\textsuperscript{$\dagger$}}{1.29} &
\metricpm{16.66\textsuperscript{$\dagger$}}{0.90} &
\metricpm{0.33\textsuperscript{$\dagger$}}{0.006} &
\metricpm{2.97\textsuperscript{$\dagger$}}{0.03} &
\metric{--} &
\metric{0.10} \\

C-coupling + PFG &
232M & 60K EN &
\metricpm{\underline{18.38}\textsuperscript{$\dagger$}}{0.85} &
\metricpm{\underline{8.96}\textsuperscript{$\dagger$}}{0.46} &
\metricpm{\underline{0.35}\textsuperscript{$\dagger$}}{0.007} &
\metricpm{\underline{3.17}\textsuperscript{$\dagger$}}{0.03} &
\metric{--} &
\metric{0.13} \\

C-coupling + PFG + SC-ReMask &
232M & 60K EN &
\metricpm{\textbf{8.39}\textsuperscript{$\dagger$}}{0.55} &
\metricpm{\textbf{3.56}\textsuperscript{$\dagger$}}{0.25} &
\metricpm{\textbf{0.42}\textsuperscript{$\dagger$}}{0.006} &
\metricpm{\textbf{3.77}\textsuperscript{$\dagger$}}{0.02} &
\metricpm{3.46}{0.34} &
\metric{0.10} \\
\bottomrule
\end{tabular}
\end{table*}

The rate $r^{\mathrm{rm}}(t_k)$ is added only to eligible generated suffix positions, excluding the acoustic prompt and positions that are already masked. Thus, SC-ReMask does not alter the pinned prompt, but allows previously generated suffix tokens to return to the mask state and be regenerated in later tau-leaping steps. Unless stated otherwise, all main experiments use $t_{\mathrm{switch}}=0$, $\eta_{\mathrm{rescale}}=0.5$, and $\eta_{\mathrm{cap}}=0.5$. These values are selected on the LibriSpeech~\cite{panayotov2015librispeech} dev-clean subset, with the schedule ablation reported in Fig.~\ref{fig:remask_ablation}.

\section{Methods}
\label{sec:methods}

We train on the English portion of Emilia-YODAS from the Emilia dataset family~\cite{he2024emilia}. We choose Emilia-YODAS to ease reproducibility and artifact sharing, since it is released under CC BY 4.0, whereas the original Emilia split is CC BY-NC 4.0, which introduces non-commercial restrictions that can complicate redistribution and reuse.

We use NeuCodec~\cite{julian2025finite}, an FSQ-based neural audio codec derived from XCodec2~\cite{ye2025llasa}, as the discrete acoustic representation because it combines low-bitrate speech reconstruction with a single-codebook tokenization scheme. This choice is particularly suitable for our CTMC formulation, since it keeps speech as a single discrete sequence, avoiding the multi-stream prediction problem of RVQ-based codecs, while recent single-stream and low-bitrate tokenizers have shown strong performance in zero-shot TTS and speech generation~\cite{guo2025recent, wang2025spark, xie2025fireredtts, wang2026tadicodec}. Text is tokenized with a GPT-2 Byte Pair Encoding (BPE) tokenizer~\cite{radford2019language}.

For evaluation, we follow the voice-prompted protocol in~\cite{wang2023neural} on LibriSpeech test-clean~\cite{panayotov2015librispeech}, using a 2.2-hour subset with utterance durations between 4 and 10 seconds. For each target utterance, we sample a random utterance from the same speaker as the prompt. All systems use the same prompt selection and preprocessing.

G-DFlow-TTS totals 232M parameters, with a DiT backbone comprising 12 layers, 12 attention heads, hidden size 768, and Rotary Position Embeddings (RoPE)~\cite{su2024roformer} with $\theta=10000$. Training runs for 1M iterations on a single NVIDIA B200 GPU with AdamW~\cite{loshchilov2018decoupled}, learning rate $3\times 10^{-4}$, cosine decay with 5{,}000 warmup steps, weight decay $1\times 10^{-6}$, effective batch size of 64 (16 per-device with 4 gradient accumulation steps), gradient clipping at 10, and mixed precision.  To enable predictor-free guidance (PFG) at inference time~\cite{nisonoff2025unlocking}, we train a single model to produce both conditional and unconditional predictions via text dropout: for $10\%$ of training examples, we replace the text condition with a filler sequence while keeping the acoustic target unchanged. At inference, we obtain the conditional and unconditional CTMC rate predictions from the same checkpoint by respectively providing the text input or the text filler input.



Intelligibility is measured by Word Error Rate (WER) and Character Error Rate (CER), computed from transcriptions produced by a CTC-based HuBERT ASR model~\cite{hsu2021hubert}. Speaker similarity (SIM-o) is the cosine similarity between WavLM-TDCNN~\cite{chen2022wavlm} embeddings of generated and reference audio. Perceptual quality is assessed with UTMOS~\cite{saeki22c_interspeech} as an objective MOS predictor, and a 5-point Likert MOS study with 19 listeners on 20 clips per system (20 distinct speakers, 10 male/10 female), randomly sampled from LibriSpeech test-clean. For per-utterance metrics, we report mean paired differences with 95\% bootstrap confidence intervals (10{,}000 resamples) and $p$-values from paired sign-flip permutation tests (10{,}000 permutations). For MOS, we bootstrap at the clip level and additionally use paired tests on clip-level means.

\begin{table*}[ht]
\centering
\scriptsize
\setlength{\tabcolsep}{3pt}
\renewcommand{\arraystretch}{1.05}
\caption{Paired significance at NFE=32 on the same utterances. $\Delta$ is the mean paired difference (variant$-$baseline). For WER/CER (in percentage points), negative is better; for SIM-o/UTMOS, positive is better. 95\% CI: paired bootstrap (10k). $p$: paired sign-flip permutation (10k); all entries have $p<10^{-4}$. \textit{Takeaway: PFG yields large gains over the baseline, while C-coupling and SC-ReMask provide additional consistent improvements.}}
\label{tab:significance}
\begin{tabular}{@{}L{6cm}cccc@{}}
\toprule
\textbf{Comparison G-DFlow-TTS (NFE=32)} &
\textbf{$\Delta$WER (\%)} &
\textbf{$\Delta$CER (\%)} &
\textbf{$\Delta$SIM-o} &
\textbf{$\Delta$UTMOS} \\
\midrule
\multicolumn{5}{@{}l}{\textbf{(A) Versus U-Coupling Baseline}} \\
U-Coupling + PFG &
\dciinline{-46.83}{-48.51,\,-45.24} &
\dciinline{-30.59}{-31.72,\,-29.46} &
\dciinline{+0.16}{+0.16,\,+0.17} &
\dciinline{+0.85}{+0.82,\,+0.88} \\

C-coupling + PFG &
\dciinline{-57.06}{-58.67,\,-55.48} &
\dciinline{-38.30}{-39.39,\,-37.24} &
\dciinline{+0.18}{+0.17,\,+0.19} &
\dciinline{+1.05}{+1.02,\,+1.08} \\

C-coupling + PFG + SC-ReMask &
\dciinline{-67.05}{-68.58,\,-65.56} &
\dciinline{-43.70}{-44.77,\,-42.67} &
\dciinline{+0.25}{+0.24,\,+0.25} &
\dciinline{+1.65}{+1.63,\,+1.68} \\
\midrule
\multicolumn{5}{@{}l}{\textbf{(B) Incremental ablations}} \\
C-coupling + PFG \textbf{vs} U-Coupling + PFG &
\dciinline{-10.23}{-11.51,\,-8.94} &
\dciinline{-7.71}{-8.60,\,-6.83} &
\dciinline{+0.02}{+0.01,\,+0.02} &
\dciinline{+0.20}{+0.17,\,+0.23} \\

C-coupling + PFG + SC-ReMask \textbf{vs} U-Coupling + PFG &
\dciinline{-20.22}{-21.45,\,-19.01} &
\dciinline{-13.11}{-13.99,\,-12.25} &
\dciinline{+0.08}{+0.08,\,+0.09} &
\dciinline{+0.80}{+0.78,\,+0.83} \\

C-coupling + PFG + SC-ReMask \textbf{vs} C-coupling + PFG &
\dciinline{-9.99}{-10.81,\,-9.17} &
\dciinline{-5.40}{-5.84,\,-4.96} &
\dciinline{+0.07}{+0.06,\,+0.07} &
\dciinline{+0.60}{+0.58,\,+0.63} \\
\bottomrule
\end{tabular}
\end{table*}

\section{Results and Discussion}
\label{sec:results}
\begin{figure}[ht]
    \centering
    \includegraphics[width=0.45\textwidth]{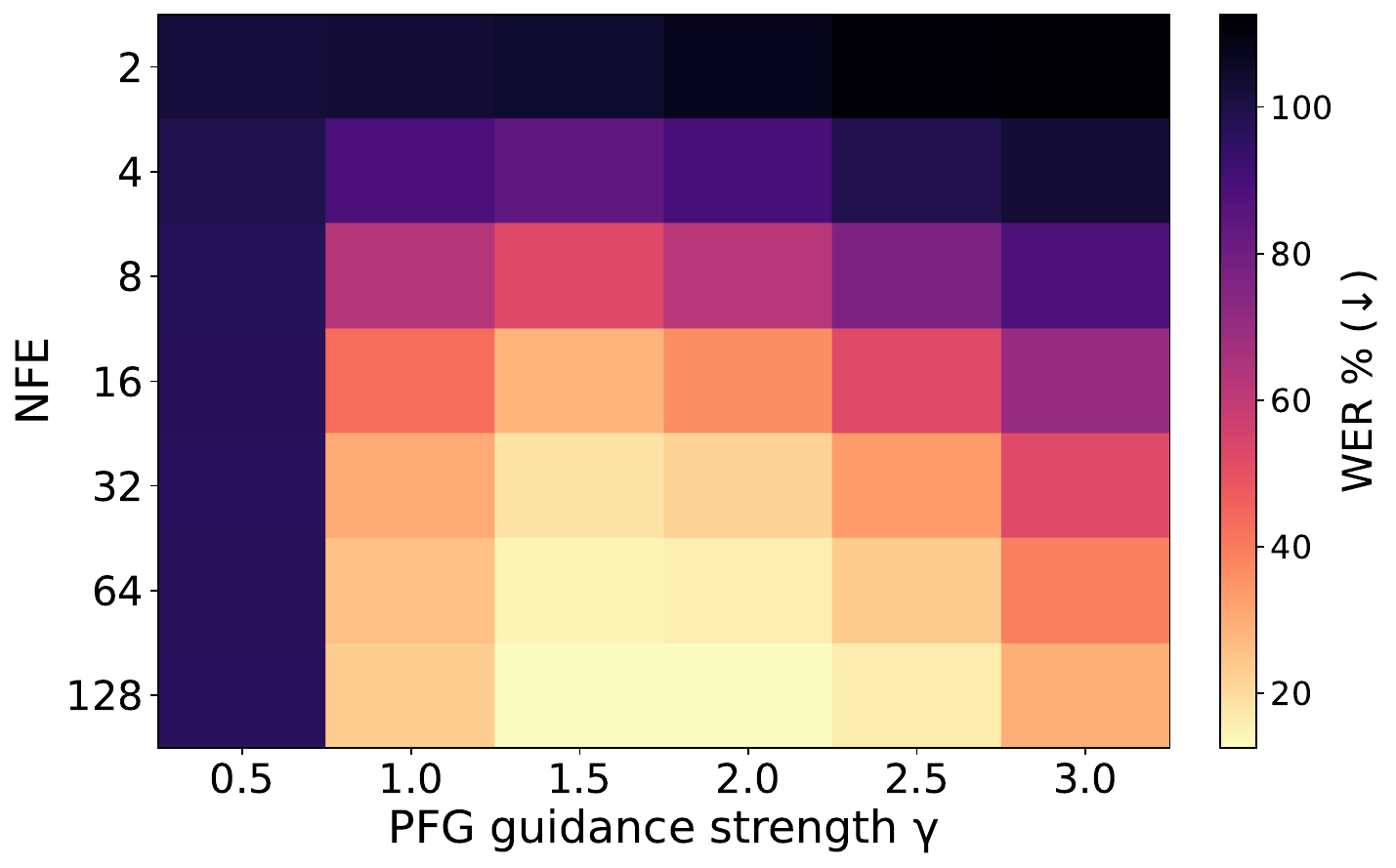}
    \caption{Grid search over predictor-free guidance strength $\gamma$ and sampling budget (NFE) for G-DFlow-TTS with conditional coupling. Colors denote WER on the LibriSpeech test-clean prompted subset (lower is better).}
    \label{fig:heatmap_grid_search}
\end{figure}

\noindent\textbf{PFG is necessary for conditional control at low NFE.}
We first tune predictor-free guidance (PFG) by sweeping the guidance strength $\gamma$ and the sampling budget (NFE). Figure~\ref{fig:heatmap_grid_search} shows that stronger guidance substantially improves intelligibility at low NFEs, but overly large $\gamma$ can degrade WER, especially when the sampling budget is small. Across this grid, the best operating point occurs at a moderate guidance strength ($\gamma = 1.5$), and we use this configuration for the remaining experiments.

\begin{figure}[!ht]
    \centering
    \includegraphics[width=0.45\textwidth]{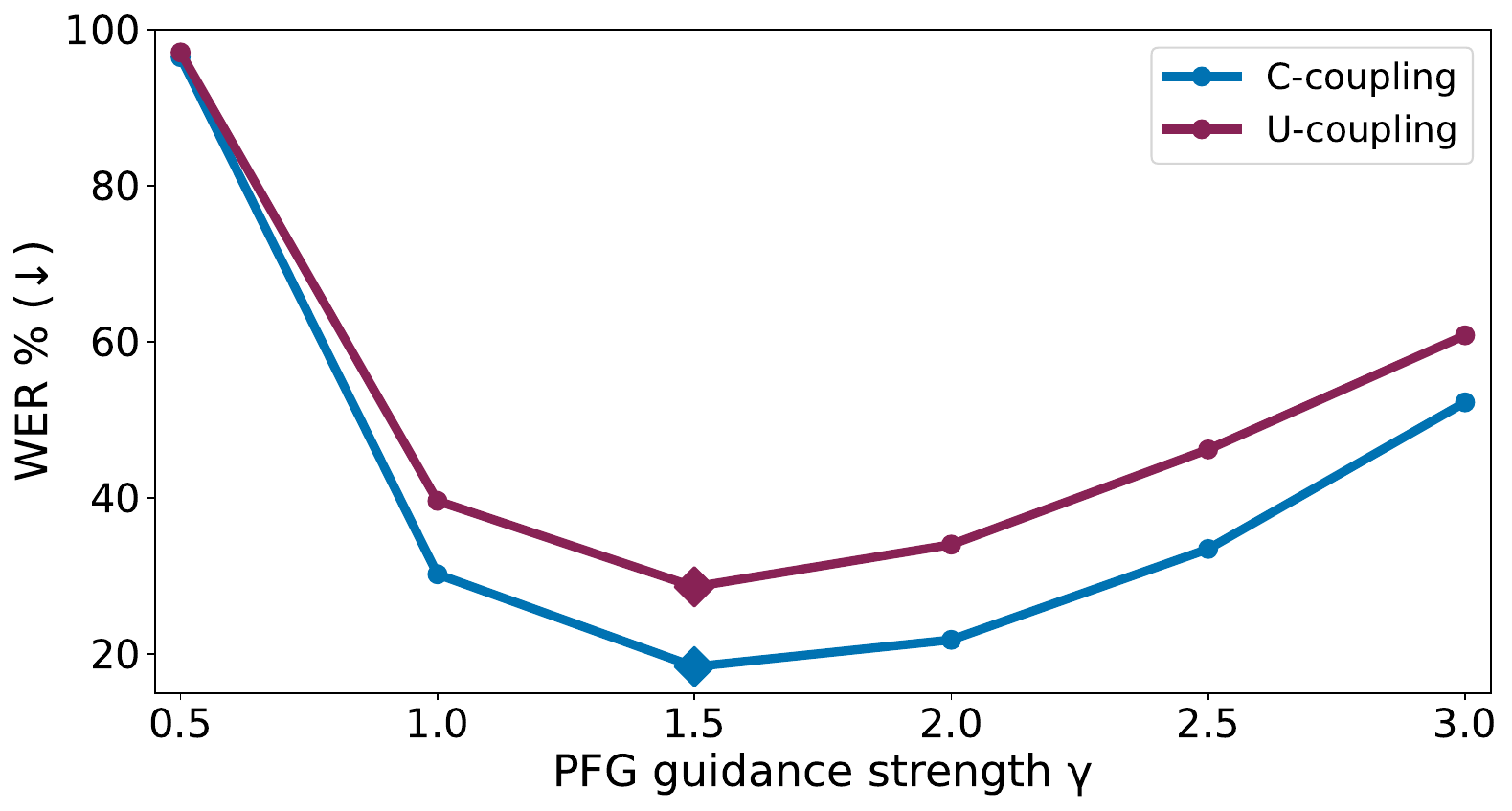}
    \caption{Effect of predictor-free guidance strength $\gamma$ at fixed sampling budget (NFE$=32$). Conditional coupling consistently improves intelligibility over U-coupling across $\gamma$. Diamonds indicate the best $\gamma$ for each coupling strategy.}
    \label{fig:pfg_grid_search}
\end{figure}

\noindent\textbf{Conditional coupling helps when paired with guided sampling.}
Figure~\ref{fig:pfg_grid_search} isolates the effect of $\gamma$ at NFE$=32$ and compares unconditional coupling (U-coupling) to conditional coupling (C-coupling). C-coupling consistently improves intelligibility across a wide range of $\gamma$, indicating that exposing the model to prompted infilling during training better matches the conditional generation task at inference. However, C-coupling alone degrades performance (Table~\ref{tab:main_results_new}), showing that prompt-matched probability paths are not sufficient without sampling-time control. This suggests that coupling and guidance play complementary roles: C-coupling makes the training path match the prompted task, while PFG strengthens the conditional CTMC rates during sampling.

\begin{figure*}[!h]
    \centering
    \includegraphics[width=\textwidth]{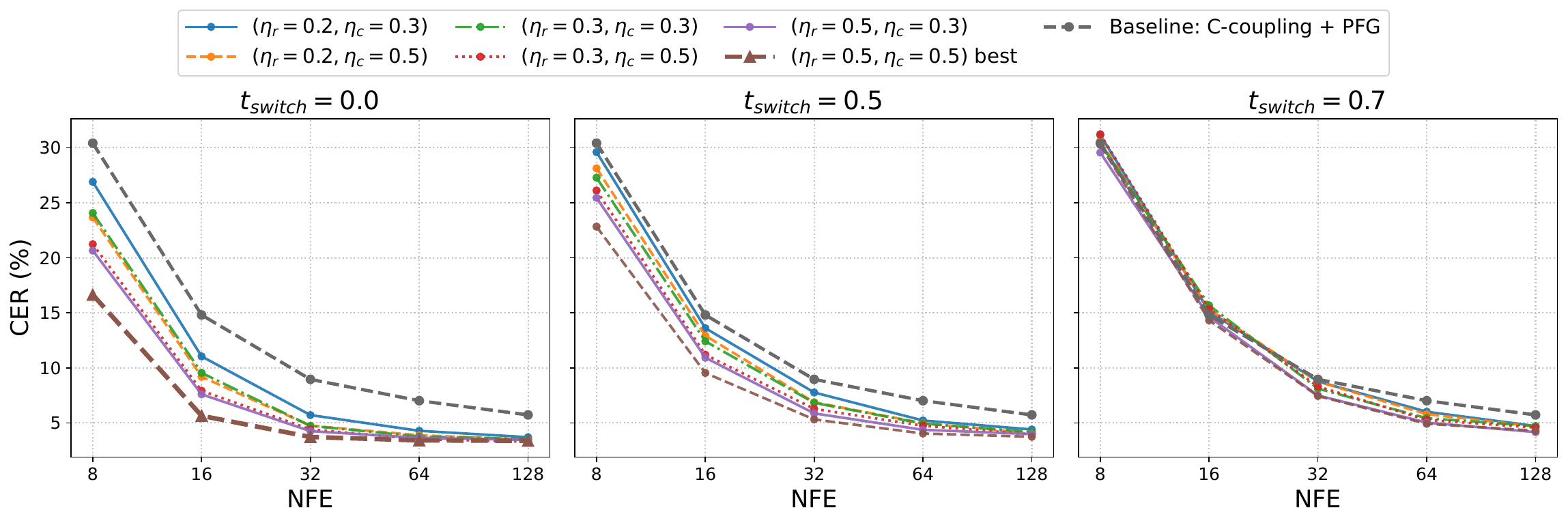}
    \caption{Schedule-constrained remasking ablation on LibriSpeech dev-clean, where $\eta_{r} = \eta_{rescale}$ and $\eta_{c}
     = \eta_{cap}$. We report CER as a function of the number of function evaluations (NFE) while varying the remasking switch time $t_{\mathrm{switch}}$, rescaling factor $\eta_r$, and cap $\eta_c$. The no-remasking baseline uses C-coupling with PFG. Always-on remasking ($t_{\mathrm{switch}}=0$) with $(\eta_r,\eta_c)=(0.5,0.5)$ gives the best low-NFE behavior among the tested schedules, showing that revisable token-to-$\MASK$ transitions improve CTMC sampling beyond guided unmasking alone.}
    \label{fig:remask_ablation}
\end{figure*}

\noindent\textbf{SC-ReMask benefits from always-on schedule-constrained revision.}
Figure~\ref{fig:remask_ablation} reports the dev-clean ablation used to select the SC-ReMask schedule. We vary the switch time $t_{\mathrm{switch}}$, rescaling factor $\eta_{\mathrm{rescale}}$, cap $\eta_{\mathrm{cap}}$, and sampling budget. Among the tested schedules, always-on remasking with $t_{\mathrm{switch}}=0$ and $(\eta_{\mathrm{rescale}},\eta_{\mathrm{cap}})=(0.5,0.5)$ gives the best low-NFE behavior. This indicates that revisable token-to-mask transitions are most useful when available throughout sampling, while the schedule constraint prevents excessive remasking. We therefore use this configuration in the main experiments.

\noindent\textbf{SC-ReMask makes generation revisable and improves robustness.}
Table~\ref{tab:main_results_new} includes external reference systems and controlled G-DFlow-TTS ablations. The external systems are reported only for contextual reference using official checkpoints, whereas the main scientific comparison is the controlled ablation within G-DFlow-TTS under a fixed backbone, training setup, prompt protocol, and evaluation pipeline. The unguided G-DFlow-TTS baseline fails in the prompted low-NFE setting (WER 75.44\%), and C-coupling alone further degrades WER, confirming the brittleness of unguided conditional infilling. In contrast, PFG is the main source of conditional control: at the same NFE, it reduces WER to 28.61\%. Combining C-coupling with PFG further reduces WER to 18.38\%, and adding SC-ReMask produces the best overall configuration, with WER 8.39\% and CER 3.56\%. SC-ReMask also improves SIM-o and UTMOS, suggesting that revising early token decisions improves not only intelligibility but also broader synthesis robustness. All reported gains over the baseline at NFE$=32$ are statistically significant under paired tests, and Table~\ref{tab:significance} shows that each component contributes incremental, significant improvements.

\noindent\textbf{Inference-time control matters more than additional sampling steps.}
Table~\ref{tab:tradeoff} examines the quality-speed trade-off. The unguided baseline saturates quickly: increasing NFE from 32 to 128 reduces CER only from 47.25\% to 40.39\%, suggesting that the main bottleneck is not simply sampling resolution. In contrast, the full Mask, Sample, Revise stack at only 8 NFE already reaches 15.92\% CER, outperforming the unguided baseline at 128 NFE. This gives strong evidence that guided and revisable inference might be the primary driver of content accuracy in the prompted low-NFE regime. Increasing NFE further improves the full system, but the largest gains come from the inference-time control stack itself rather than from more CTMC steps alone.


\noindent\textbf{Limitations.}
Speaker similarity remains below large external baselines (Table~\ref{tab:main_results_new}), likely because our model relies only on an acoustic prefix and does not use an explicit speaker objective. As a result, timbre may drift when early errors propagate. We also note that SIM-o is an embedding-space proxy that may not fully reflect perceived identity~\cite{ahn24b_interspeech}, and codec artifacts can affect speaker embeddings~\cite{ferrofilho25_interspeech}. Finally, our Emilia-YODAS setup does not apply the transcription-quality or language filters used in systems such as F5-TTS~\cite{chen-etal-2025-f5}, which may partially explain the gap to stronger filtered baselines.


\begin{table}[!t]
\centering
\caption{Quality-speed trade-off across sampling budgets evaluated on LibriSpeech test-clean. In the \textit{System} column, \textit{Baseline} denotes the G-DFlow-TTS baseline using U-coupling, \textit{PFG + CC} denotes PFG with C-coupling, and \textit{Full} denotes our complete method combining C-coupling, PFG, and SC-ReMask.}

\label{tab:tradeoff}
\setlength{\tabcolsep}{4pt}
\sisetup{
    output-decimal-marker={.},
    round-mode=places,
    round-precision=2
}
\begin{tabular}{
    c
    l
    S[table-format=1.3]
    S[table-format=1.3]
    S[table-format=1.3]
    S[table-format=1.3]
}
\toprule
\thead{NFE} & \thead{System} & {\thead{CER (\%) $\downarrow$}} & {\thead{SIM-o $\uparrow$}} & {\thead{UTMOS $\uparrow$}} & {\thead{RTF $\downarrow$}} \\
\midrule
4  & Baseline & 83.05 & 0.10 & 1.45 & 0.01 \\
4  & PFG + CC & 53.04 & 0.18 & 1.85 & 0.02 \\
4  & Full  & 43.95 & 0.21 & 2.14 & 0.01 \\
\midrule
8  & Baseline & 68.30 & 0.13 & 1.70 & 0.01 \\
8  & PFG + CC & 30.42 & 0.25 & 2.35 & 0.03 \\
8  & Full  & 15.92 & 0.328 & 3.02 & 0.03 \\
\midrule
16 & Baseline & 55.18 & 0.15 & 1.96 & 0.03 \\
16  & PFG + CC & 14.81 & 0.31 & 2.85 & 0.06 \\
16 & Full  & 5.69 & 0.398 & 3.57 & 0.05 \\
\midrule
32 & Baseline & 47.25 & 0.17 & 2.12 & 0.05 \\
32  & PFG + CC & 8.96 & 0.347 & 3.17 & 0.13 \\
32 & Full  & 3.56 & 0.415 & 3.77 & 0.10 \\
\midrule
64 & Baseline & 43.17 & 0.18 & 2.23 & 0.10 \\
64  & PFG + CC & 7.02 & 0.365 & 3.308 & 0.25 \\
64 & Full  & 3.22 & 0.412 & 3.81 & 0.21 \\
\midrule
128 & Baseline & 40.39 & 0.18 & 2.26 & 0.20 \\
128  & PFG + CC & 5.73 & 0.378 & 3.40 & 0.50 \\
128 & Full  & 3.00 & 0.411 & 3.82 & 0.41 \\
\bottomrule
\end{tabular}
\end{table}

\section{Conclusion}
\label{sec:conclusion}

We presented G-DFlow-TTS, an alignment-free codec-token TTS system built on Discrete Flow Matching and CTMC sampling. Our results show that stable conditional infilling in DFM-based TTS depends critically on inference-time control. The proposed Mask, Sample, Revise stack combines predictor-free guidance, prompt-matched conditional coupling, and SC-ReMask within a single tau-leaping sampler. Across objective metrics and human listening tests, these components significantly improve intelligibility and robustness, with SC-ReMask providing the largest gains by allowing early token decisions to be revised. Future work will explore stronger speaker conditioning, better-filtered training data, and improved token representations to reduce the gap to high-resource systems while preserving alignment-free sampling.


\section*{Acknowledgments} 

This work has been funded by the project Research and Development of Genese Digital: Scaling Interactive and Culturally Adapted Digital Humans with Generative AI, supported by the Advanced Knowledge Center in Immersive Technologies (AKCIT), with financial resources from the PPI IoT/Manufatura 4.0 / PPI HardwareBR of the MCTI, grant number 057/2023, signed with EMBRAPII.

\section*{Generative AI Use Disclosure}

The authors used generative AI tools, specifically ChatGPT (OpenAI, GPT-5.5 Thinking), only to polish portions of the manuscript. These tools were used only for minor editing tasks, such as improving grammar, clarity, and phrasing. They were not used to generate scientific ideas, design experiments, analyze results, generate figures, generate code used in the experiments, or write major parts of the paper. The authors carefully reviewed and edited all text and take full responsibility for the final content of the manuscript.



\bibliographystyle{IEEEtran}
\bibliography{refs}
\end{document}